\documentclass[aps,twocolumn,nofootinbib]{revtex4}
\usepackage[latin1]{inputenc}
\usepackage{epsfig}
\newcommand{\beq}{\begin{equation}}
\newcommand{\eeq}{\end{equation}}
\newcommand{\la}{\langle}
\newcommand{\ra}{\rangle}

\begin{document}

\title{Stochastic interpretation of quantum mechanics}
\author{Mário J. de Oliveira}
\affiliation{Universidade de São Paulo,
Instituto de Física,
Rua do Matão, 1371, 05508-090
São Paulo, SP, Brasil}

\begin{abstract}

We express the probabilistic character associated to the
wave function by treating it as a stochastic variable.
This is accomplished by means of a stochastic equation
for the wave function whose noise changes the phase of
the wave function but not its absolute value, so that the
norm of the wave function is strictly conserved along a
stochastic trajectory. We show that the density matrix
that obeys the quantum Liouville equation is the covariance
matrix associated to the stochastic wave function.

\end{abstract}

\maketitle

\section{Introduction}

Quantum mechanics \cite{landau1958,merzbacher1961,messiah1961,%
sakurai1967,sakurai1994,griffiths1995,piza2002,griffiths2002}
differs fundamentally from classical mechanics
\cite{lanczos1949,goldstein1950,landau1960,arnold1978}.
In contrast to the latter, the former contains
quantities that do not have obvious interpretation.
In classical mechanics the state of a system is defined by 
positions an momenta of the particles, which are
interpreted as the real positions and the real momenta
of the particles. In quantum mechanics, on the other hand,
the state is defined by the wave function, which has
no obvious interpretation.

The wave function is not observed experimentally and
for that reason it is called unobservable. For many,
this terminology implies, usually in a tacit manner, 
a metaphysical conception of the wave function. That is,
a concept that is real but which is over and beyond all
experience \cite{carnap1935}. 
However, a reality which is beyond experience
is inconsistent and contradictory, at least in the 
realm of science. If we deny such an understanding we
may be lead to propositions that are devoid of scientific
meaning or even unscientific.
In fact, there is no need to assign a
real existence of an unobservable if we understand a
scientific theory as a representation of the real and
not the real itself. 

In accordance with the standard interpretation of quantum
mechanics \cite{omnes1994,auletta2001,barandes2020,%
freire2022,barioni2022,pessoa2025}
the square of the absolute value of the wave function
is related to probability. More precisely, if $\psi(x)$ 
is the wave function associate to the position $x$ 
of a particle that moves along
a straight line then $|\psi(x)|dx$ is the probability of
finding the particle between $x$ and $x+dx$.
However, we need also to state the meaning or the
interpretation of probability. Within the standard 
interpretation, it seems that it is interpreted
as related to frequency, which can be measured
experimentally.

In accordance with the standard interpretation, the
probabilistic character of the wave function emerges at
the instant of measurement, when the wave function is
said to collapse. A question then arises as to the
probabilistic character of the wave function prior to
measurement. That is, suppose that we perform a
measurement at time $t$ and ask whether at a time
$t'<t$ the wave function has the probabilistic character.
To answer this question we must take a measurement at the
instant $t'$. But, if we do that the wave function collapses
at $t'$ and the measurement at time $t$ becomes compromised.

From these reasonings, we may say that the probabilistic
character of the wave function prior to measurement is not
liable to measurement. Within the standard interpretation of
quantum mechanics, it is assumed that the wave function
is not associated to probability prior to measurement,
and that the probabilistic character emerge at the instant
of measurement. Our aim here is to provide an approach
in which the wave function does have a probabilistic
character at any time. This is accomplished by treating
the wave function as a stochastic variable which obeys
a stochastic equation of motion 
\cite{oliveira2024a,oliveira2024b,oliveira2025}.
It is important to say that the probability character
at the instant of measurement given by the present
approach is the same as that given by the standard
interpretation. 

Our stochastic approach should not be confused or identified
with previous stochastic approaches \cite{nelson1966} in which
the stochastic variable is the quantum position of the particle
and not the wave function. Notice that in our approach the
number of stochastic variable is proportional to the
dimension of the Hilbert space whereas in these previous
stochastic approaches it is equal to the number of degrees
of freedom of the quantum system, which is one for
a quantum particle moving on a straight line.

From our approach it follows that the covariance matrix
associated to the stochastic wave function is identified
as the quantum density matrix.
The equation that one obtains for the covariances,
which is a closed equation in the covariances,
is then identified with the quantum Liouville equation.
This equation may have solutions consisting of the
product of a function and its complex conjugate.
When this happens these functions are identified
as the pure states. 

The stochastic equations that we use are the 
classical Hamiltonian equations of motion complemented
by a noise. The noise we use is a special type of
noise that preserves the norm of the stochastic
wave function. This is not merely a technical detail 
but has a conceptual significance
because without it we would not obtain quantization 
properties from classical motion. 

We are interested here only in the part of the
interpretation of quantum mechanics associated with the
probabilistic aspect considering our stochastic
treatment of the wave function. This treatment
allows to understand that the probabilistic
character of quantum mechanics occurs at all times,
and not just at the time of measurement, thus making
the collapse of the wave function unnecessary.

\section{Stochastic equations of motion}

We consider a state vector $|\psi\ra$ defined on
a Hilbert space of dimension $n$,
\beq
|\psi\ra = \sum_j z_j |j\ra,
\label{3}
\eeq
where $z_j$ are the components of the state vector.
The time evolution is given by the Schrödinger
equation
\beq
\frac{d}{dt}|\psi\ra = \frac1{ih}H|\psi\ra,
\eeq
where $h$ is the Planck constant and
$H$ is a $n\times n$ Hermitian matrix,
the Hamiltonian matrix.
This equation can also be written in terms
of the components $z_j$. Replacing (\ref{3}) 
into this equation we find
\beq
\frac{dz_j}{dt} =  \frac1{ih} \sum_k H_{jk} z_k,
\label{15a}
\eeq
where $H_{jk}$ are the elements of the matrix $H$,
which holds the Hermitian property $H_{jk}^*=H_{kj}$.

If we discretize the
time in equal interval of time $\tau$ and
denoting by $|\psi'\ra$ the state vector at
the instant $t+\tau$, then we may write
\beq
|\psi'\ra = |\psi\ra + \frac\tau{ih}H|\psi\ra.
\label{11}
\eeq

Next we turn $z_j$ into stochastic variable
that obey a stochastic equation which we
establish by adding a noise into equation 
(\ref{11}). We assume the following stochastic
equation
\beq
|\psi'\ra = e^{i\sqrt{\tau} \xi} |\psi\ra
+ \frac\tau{ih} H| \psi\ra,
\label{13}
\eeq
where $\xi$ is a random variable with zero mean
and variance $\overline{\xi^2}=\gamma$.
It is clear from this equation that the noise
changes the phase of the state vector $|\psi\ra$
but not its absolute value.

From equation
(\ref{13}), we find
\beq
\la\psi'|\psi'\ra - \la\psi|\psi\ra = o(\tau),
\eeq
where $o(\tau)$ stands for terms of order
strictly lower than $\tau$. Dividing by
$\tau$ and taking the limit $\tau\to0$, we find
\beq
\frac{d}{dt}\la\psi|\psi\ra = 0,
\eeq
which means that the norm is strictly conserved along
the {\it stochastic trajectory} of the state vector
in the Hilbert space. From now on we set the norm
equal to one, that is,
\beq
\la\psi|\psi\ra = \sum_j z_j z_j^* = 1.
\eeq

The equation (\ref{13}) can be written in terms of the components
$z_j$ of the state vector. If we let $z_j'$ be the
component $z_j$ at the instant $t+\tau$ then
\beq
z_j' = e^{i\sqrt{\tau} \xi} z_j
+  \frac\tau{ih} \sum_k H_{jk} z_k.
\eeq
Up to terms of order $\tau$ this equation can be written as
\beq
z_j' = (1+ i\sqrt{\tau} \xi -\frac\tau2 \xi^2) z_j
+  \frac\tau{ih} \sum_k H_{jk} z_k.
\label{15}
\eeq
Defining the increment $\Delta z_j=z_j'-z_j$, we may
write
\beq
\Delta z_j =  (i\sqrt{\tau} \xi -\frac\tau2 \xi^2) z_j
+  \frac\tau{ih} \sum_k H_{jk} z_k.
\label{16}
\eeq

\section{Underlying classical system}

We show here that the equations (\ref{15a}) are equivalent
to the Hamilton equations of motion of a classical system
of $n$ degrees of freedom, which we call the underlying
system. We denote by $q_j$ and $p_j$ the coordinate and
the momentum associated to a degree of freedom. They 
form a pair of canonically conjugate variables, and
the Hamilton equations of motion are 
\beq
\frac{dq_j}{dt} = \frac{\partial{\cal H}}{\partial p_j},
\qquad \frac{dp_j}{dt} = -\frac{\partial{\cal H}}{\partial q_j}
\eeq
where $\cal H$ is a function of $q_j$ and $p_j$.
We introduce now a transformation of variables from
$(q_j,p_j)$ to $(z_j,z_j^*)$ defined by
\beq
z_j = \alpha q_j + i \beta p_j
\qquad z_j^* = \alpha q_j - i \beta p_j
\eeq
where $z_j$ and its complex conjugate $z_j^*$ are understood
as independent variables, and $\alpha$ and $\beta$ are
constant to be chosen. Using this transformation,
and choosing $\alpha$ and $\beta$ such that
$2\alpha\beta=1/h$, we find the equations 
\beq
\frac{dz_j}{dt}
= \frac1{ih}\frac{\partial{\cal H}}{\partial z_j^*}
\qquad \frac{dz_j^*}{dt}
= -\frac1{ih} \frac{\partial{\cal H}}{\partial z_j}
\label{20}
\eeq
where ${\cal H}$ is understood as a function of $z_j$ and
$z_j^*$ obtained by the transformation above, and
we see that $z_j$ and $z_j^*$ can be understood as a pair
of canonically conjugate variables.

To end our demonstration, it is left to choose an 
appropriate Hamiltonian function $\cal H$ that lead
us to the equation (\ref{15a}).
We choose $\cal H$ to be of the form
\beq
{\cal H} = \sum_{jk} H_{jk}z_k z_j^*
\eeq
Replacing this form in the first of the equations (\ref{20})
we reach (\ref{15a}). If we replace it in the second of the
equations (\ref{20}) we reach an equation which is the
complex conjugate of (\ref{15a}). 

The introduction of the underlying system allows to 
interpret the real part and the imaginary parts
of the component $z_j$ of the state vector as proportional
to the coordinate $q_j$ and momentum $p_j$ associated
to a degree of freedom. We remark that the underlying
system is not to be confused or identified with the 
real quantum system and in this sense 
it could be called fictitious. But then we should
also call the wave function fictitious. Therefore,
it is more appropriate to call the underlying system
an unobservable as much as the wave function is an
unobservable.

\section{Fokker-Planck equation}

From the stochastic equation of motion
we may determine an equation for the time
evolution of the probability distribution
${\cal P}(z,z^*)$
of the set of variables $z_j$ and $z_j^*$.
To this end we start by considering a
bilinear function 
\beq
{\cal F} = \sum_{jk} F_{jk}z_j z_k^*.
\eeq
The variation of this function between to 
instants of time $t+\tau$ and $t$ is
given by
\[
\Delta{\cal F} = \sum_j (\frac{\partial{\cal F}}{\partial z_j}
\Delta z_j + \frac{\partial{\cal F}}{\partial z_j^*}
\Delta z_j^*)
\]
\beq
+ \sum_{jk} \frac{\partial^2{\cal F}}{\partial z_j\partial z_k^*}
\Delta z_j \Delta z_k^*.
\eeq
Using (\ref{15}) and the abbreviation
\beq
f_j = \frac1{ih} \sum_k H_{jk} z_k,
\label{18}
\eeq
then we find up to terms of order $\tau$
\[
\Delta{\cal F} = \sum_j \frac{\partial{\cal F}}{\partial z_j}
(\tau f_j -\frac{\sqrt\tau}2 \xi z_j -\frac\tau2 \xi^2 z_j) +
\]
\[
+ \sum_j \frac{\partial{\cal F}}{\partial z_j^*}
(\tau f_j^*-\frac{\sqrt\tau}2 \xi z_j^* -\frac\tau2 \xi^2 z_j^*) +
\]
\beq
+ \tau \xi^2\sum_{jk} \frac{\partial^2{\cal F}}{\partial z_j\partial z_k^*}
z_j z_k^*.
\eeq
From which follows 
\[
\frac{d\la{\cal F}\ra}{d t}
= \sum_j \la f_j\frac{\partial{\cal F}}{\partial z_j}
+  f_j^*\frac{\partial{\cal F}}{\partial z_j^*}\ra
\]
\beq
- \frac\gamma2 \sum_j \la z_j\frac{\partial{\cal F}}{\partial z_j}
+ z_j^*\frac{\partial{\cal F}}{\partial z_j^*}\ra
+ \gamma\sum_{jk} \la\frac{\partial^2{\cal F}}
{\partial z_j\partial z_k^*} z_j z_k^*\ra,
\label{17}
\eeq
where the averages are obtained using the probability
distribution ${\cal P}(z,z^*)$.

The equation that gives the time evolution of the
probability distribution is obtained from (\ref{17})
and is given by
\[
\frac{\partial \cal P}{\partial t}
= - \sum_j (\frac{\partial f_j {\cal P}}{\partial z_j}
+\frac{\partial f_j^*{\cal P}}{\partial z_j^*})
\]
\beq
+\frac\gamma2 \sum_j (\frac{\partial z_j {\cal P}}{\partial z_j}
+ \frac{\partial z_j^*{\cal P}}{\partial z_j^*})
+ \gamma\sum_{jk} \frac{\partial^2 z_j z_k^*{\cal P}}
{\partial z_j\partial z_k^*}.
\label{50}
\eeq
which is a Fokker-Planck equation 
\cite{vankampen1981,tome2015}.

\section{Covariances}

From the probability distribution ${\cal P}$, we may determine
$\la z_j\ra$, and the covariances $\la z_jz_k^*\ra$.
The time evolution of these quantities can be obtained
from the equation (\ref{17}) and they are given by
\beq
\frac{d\la z_j \ra}{d t} = \la f_j \ra -\frac\gamma2 \la z_j\ra,
\eeq
and
\beq
\frac{d\la z_j z_k^* \ra}{d t}
= \la f_j z_k^* +  f_k^* z_j \ra,
\eeq
and we see that the terms involving $\gamma$ disappears
from this last equation. Using (\ref{18}) we obtain
\beq
\frac{d\la z_j z_k^* \ra}{d t}
= \frac1{ih} \sum_l (H_{jl} \la z_l z_k^*\ra
- H_{lk} \la z_l^* z_j \ra).
\eeq
Defining the density matrix $\rho_{jk}$ by
\beq
\rho_{jk} = \la z_j z_k^*\ra,
\eeq
we reach the equation
\beq
\frac{d\rho_{jk}}{d t}
= \frac1{ih} \sum_l (H_{jl} \rho_{lk }
- H_{lk} \rho_{jl}),
\label{19}
\eeq
which is the quantum Liouville equation.
We observe, however, that from the derivation
of this equation that $\rho_{jk}$ is understood
as the {\it covariances} of the components of the
stochastic state vector $|\psi\ra$.
Since the norm is equal to one then
\beq
\sum_j \rho_{jj} = 1.
\eeq
Defining the $n\times n$ matrix $\rho$ whose elements
are $\rho_{jk}$ then the equation (\ref{19})
can be written in the matrix form
\beq
\frac{d\rho}{d t}
= \frac1{ih} (H\rho - \rho H).
\eeq

The Liouville equation has a special type of
solution of the type $\rho_{jk}(t) = \phi_j(t)\phi_k^*(t)$.
Indeed, if we replace this form in equation (\ref{19})
we find
\beq
\phi_j \frac{d\phi_k^*}{d t} + \phi_k^*\frac{d\phi_j}{d t}
= \frac{\phi_k^*}{ih}  \sum_l H_{jl} \phi_l 
- \frac{\phi_j}{ih} \sum_l  H_{lk}  \phi_l^*,
\eeq
and we see that this is possible if
\beq
\frac{d\phi_j}{d t}
= \frac1{ih}  \sum_l H_{jl} \phi_l,
\label{21}
\eeq
and as long as the initial condition
is such that $\rho_{jk}(0) = \phi_j(0)\phi_k^*(0)$
for all pairs $jk$. 

If we define the vector
\beq
|\psi_s\ra = \sum_j \phi_j |j\ra.
\eeq
Then using (\ref{21}) it follows that it
obeys the equation
\beq
\frac{d}{dt}|\psi_s\ra = \frac1{ih} H|\psi_s\ra,
\label{23}
\eeq
which is identified with the
Schrödinger equation. We call $|\psi_s\ra$
the Schrodinger state vector 
which should not be confused with the stochastic
state vector $|\psi\ra$ given by (\ref{3})
which obey the stochastic equation (\ref{13}).

It is worthwhile to compare the equations
obeyed by these two state vector. To this
end we write (\ref{23}) in the discrete time
form
\beq
|\psi_s'\ra = |\psi_s\ra + \frac\tau{ih} H|\psi_s\ra,
\eeq
which should be compare with (\ref{13}) which
we rewrite here
\beq
|\psi'\ra = e^{i\sqrt{\tau} \xi} |\psi\ra
+ \frac\tau{ih} H| \psi\ra,
\eeq
and we recall that $\xi$ is a random variable,
absent in the previous equation.

The relation between the components of the
the Schrödinger state vector and the stochastic
state vector is
\beq
\phi_j\phi_k^* = \la z_jz_k^*\ra,
\eeq
and in particular
\beq
|\phi_j|^2 = \la |z_j|^2\ra,
\eeq
and, according to the present approach, $|\phi_j|^2$
is not properly a probability but is a variance associated to the
stochastic state vector, and the off diagonal terms $\phi_j\phi_k^*$
are the covariances associated to the
stochastic state vector.

\section{Continuous space}

We now consider an infinite-dimensional 
Hilbert space. We assume that the wave function
$\psi(x,t)$ is given by 
\beq
\psi(x) = \sum_j c_j \phi_j(x),
\label{24}
\eeq
where $\phi_j(x)$ are the eigenfunctions of a
certain differential operator which forms a
complete set of orthonormalized functions.
The $c_j$ are coefficients which depend on time and are
understood as stochastic variables.

In an analogous way we assume the following
stochastic equation in discrete time
\beq
\psi' = e^{i\sqrt{\tau}\xi} \psi + \frac{\tau}{ih}\hat{H} \psi,
\label{25}
\eeq
where here $\hat{H}$ is the Hamiltonian differential
operator and $\psi'$ is the wave function at time 
$t+\tau$. Using the same reasoning as before, we conclude that 
the norm of the stochastic wave function is preserved at
any stochastic trajectory,
\beq
\frac{d}{dt} \int |\psi(x)|^2 = 0,
\eeq
which allows to set it equal to a constant which we choose to
be one, that is,
\beq
\int |\psi(x)|^2 = 1.
\eeq

Replacing (\ref{24}) in (\ref{25}), we find
\beq
 c_j' = e^{i\sqrt{\tau}\xi}  c_j
+ \frac{\tau}{ih} \sum_k  H_{jk} c_k,
\label{26}
\eeq
where
\beq
H_{jk} = \int \phi_j^* \hat{H} \phi_k dx,
\label{35}
\eeq
and $c_j'$ stands for $c_j$ at time $t+\tau$.

Equation (\ref{26}) becomes identical to (\ref{15})
if we formally replace $c_j$ by $z_j$. Therefore
we may draw the same conclusions. For instance,
the equation (\ref{19}) becomes
\beq
\frac{d\rho_{jk}}{d t}
= \frac1{ih} \sum_l (H_{jl} \rho_{lk }
- H_{lk} \rho_{jl}),
\label{19a}
\eeq
where here
\beq
\rho_{jk} = \la c_j c_k^*\ra,
\eeq

Defining
\beq
\rho(x,x') = \la\psi(x)\psi^*(x')\ra,
\eeq
then
\beq
\rho(x,x') = \sum_{jk} \rho_{jk} \phi_j(x) \phi_k^*(x').
\eeq
Deriving with respect to time and using (\ref{19a})
we find
\[
\frac{\partial}{\partial t} \rho(x,x') = \frac1{ih}
\int H(x,x'')\rho(x'',x') dx''
\]
\beq
- \frac1{ih} \int \rho(x,x'')H(x'',x')dx'',
\label{28}
\eeq
where
\beq
H(x,x') = 
\sum_{jk}  \phi_j(x) H_{jk}\phi_k^*(x'),
\label{32}
\eeq
\beq
\rho(x,x') = 
\sum_{jk}\phi_j (x) \rho_{jk } \phi_k^*(x').
\eeq
Equation (\ref{28}) can also be written in the form
\beq
\frac{\partial}{\partial t} \rho(x,x') = \frac1{ih}  [\hat{H}_x \rho(x,x')
- \hat{H}_{x'} \rho(x,x')],
\label{29}
\eeq
where the differential operators $\hat{H}_x$ and 
$\hat{H}_{x'}$ act on functions of $x$ and $x'$, respectively.

Let us consider solutions of the type
\beq
\rho(x,x') = \psi_s(x)\psi_s^*(x').
\label{30}
\eeq
If we replace in (\ref{29}), we see that this is
indeed a solution as long as
\beq
\frac{\partial\psi_s(x)}{\partial t}
= \frac1{ih} \hat{H}_x \psi_s(x),
\label{31}
\eeq
which identify with the Schrödinger equation.
Notice that $\psi_s(x)$ should no be confused with 
$\psi(x)$ given by (\ref{24}) which we call
stochastic wave function because the coefficients
$c_j$ are stochastic variables.
The solutions of the type (\ref{30}) allow us
to write $\rho_{jk}=\la c_jc_k^*\ra= a_ja_k^*$,
in which case the Schrödinger wave function 
is written as
\beq
\psi_s(x) = \sum_j a_j \phi_j(x).
\label{24a}
\eeq

It is worth comparing the equations obeyed by 
$\psi$ and by $\psi_s$. The equation obeyed by
the former is the equation (\ref{25}), which 
we rewrite here
\beq
\psi' = e^{i\sqrt{\tau}\xi} \psi + \frac{\tau}{ih}\hat{H} \psi.
\eeq
The equation obeyed by $\psi_s$ is the equation (\ref{31})
which in the discretize form is
\beq
\psi_s' = \psi_s + \frac\tau{ih} \hat{H}_x \psi_s,
\eeq
where $\psi_s'$ is the wave function at time $t+\tau$.

Let us look at the density $\rho(x,x) = |\psi_s(s)|^2$.
In the standard interpretation, $|\psi_s(s)|^2 dx$ is the
probability of finding the particle between $x$ and $x+dx$.
In the present approach $|\psi_s(s)|^2$ is the 
variance $\la |\psi(x)|^2\ra$ of the stochastic
wave function $\psi(x)$.  

\section{Underlying continuous classical system}

In the case of continuous space we can also introduce
an underlying classical system whose equations of motion
are equivalent to Schrödinger equation. In this
case the Hamilton equations of motion are
\beq
\frac{\partial\psi}{\partial t} = \frac1{ih}
\frac{\delta {\cal H}}{\delta\psi^*},
\qquad
\frac{\partial\psi^*}{\partial t} = -\frac1{ih}
\frac{\delta {\cal H}}{\delta\psi},
\eeq
where the Hamiltonian functional $\cal H$ is given by
\beq
{\cal H} = \int \psi^*(x)H(x,x')\psi(x')dx dx',
\label{37}
\eeq
and $H(x,x')$ is given by (\ref{32}).
Calculating the functional derivative we find
\beq
\frac{\partial\psi(x)}{\partial t} = \frac1{ih}.
\int H(x,x')\psi(x') dx'
\eeq
Taking into account the relations (\ref{32}) and 
(\ref{35}), we find
\beq
\int H(x,x')\psi(x') dx' = \hat{H} \psi(x),
\label{36}
\eeq
and we may write
\beq
\frac{\partial\psi}{\partial t} = \frac1{ih}\hat{H} \psi.
\eeq

If we replace (\ref{36}) into (\ref{37}), we get
\beq  
{\cal H} = \int \psi^*(x) \hat{H} \psi(x) dx.
\label{38}
\eeq
This is the formula for the {\it quantum average}
of a quantum operator associated to an observable,
in this case the energy. We are using
the term quantum average to distinguish it from the
true average obtained from a probability distribution.
The average of $\cal H$ in probability is given by
\beq
\la{\cal H}\ra
= \int H(x,x') \rho(x',x) dx dx'.
\eeq
If $\rho(x',x)=\psi_s(x')\psi_s(x)^*$ then
\beq
\la{\cal H}\ra
= \int \psi_s^*(x) H(x,x')\psi_s(x') dx dx'.
\eeq
Which can also be written as
\beq  
\la {\cal H}\ra = \int \psi_s^*(x) \hat{H} \psi_s(x) dx.
\eeq

\section{Position and momentum}

Let us write for the position and momentum
equations analogous to (\ref{37}) and (\ref{38}).
For the position we have
\beq
{\cal X} = \int \psi^*(x) \hat{x} \psi(x) dx,
\label{52}
\eeq
where $\hat{x}=x$, and
\beq
{\cal X} = \int \psi^*(x) X(x,x')\psi(x') dx dx',
\eeq
where
\beq
X(x,x') = x\delta(x-x') = \frac{x}{2\pi} \int e^{i(x-x')k} dk,
\eeq
and
\beq
\la{\cal X}\ra = \int X(x,x') \rho(x',x)dx dx'.
\label{42}
\eeq 
For the momentum, we have
\beq
{\cal Y} = \int \psi^*(x) \hat{p} \psi(x) dx,
\eeq
where $\hat{p}=-i\hbar\partial/\partial x$, and
\beq
{\cal Y} = \int \psi^*(x) Y(x,x')\psi(x') dx dx',
\label{43}
\eeq
where
\beq
Y(x,x') = i\hbar\frac1{2\pi} \int k e^{i(x-x')k} dk,
\eeq
and
\beq
\la{\cal Y}\ra = \int Y(x,x')\rho(x',x) dx dx'.
\eeq

The averages $\la{\cal X}\ra$ and $\la{\cal Y}\ra$ are understood
as the values obtained in the measurement of the position
and momentum of a particle. According to formulas (\ref{42}) and
(\ref{43}), they are related to the covariances of the
stochastic wave function. Let us compare with the standart
interpretation. To this end let us write ${\cal X}$ as
\beq
{\cal X} = \int \psi^*(x) x\psi(x) dx,
\eeq
and
\beq
\la {\cal X}\ra  = \int x\rho(x,x) dx.
\eeq
In the standard interpretation, $\rho(x,x)dx$ is the probability
of finding the particle between $x$ and $x+dx$, that is,
$\rho(x,x)$ is probability density distribution associated
to the variable $x$, which is thus understood as a random
variable. Here, on the other hand, $\rho(x,x)$ is the
variance of $\psi(x)$, understood as a stochastic variable,
associated to a certain probability distribution.
We did not write down the probability distribution for
the present case of continuous variable. But it is similar
to that of discrete variables, which obeys the 
Fokker-Planck equation (\ref{50}).

\section{Conclusion}

The probabilistic character of quantum mechanics was
expressed here by the introduction of a stochastic 
dynamics for the wave function. This was implemented
by assuming that the wave function is a stochastic
variable that obeys a stochastic equation. From this
equation we show that the density matrix is a 
covariance matrix. That is, the elements of the
density matrix are the covariances of the stochastic
wave functions. We used a noise that changes the
phase of the wave function but not its absolute
value. This implies that the stochastic dynamics
preserves strictly the norm of the wave function. 

We have also shown that a quantum system defined on
a Hilbert space of dimension $n$ can by described by
an underlying classical system of $n$ degrees of 
freedom. The real and imaginary parts of the wave
function are proportional to the coordinate and
the momentum associated to a certain degree of freedom.

From our results, the wave function function follows
a stochastic trajectory in phase space so that at
each time we may associate a probabilistic character
of the quantum system and not only at the instant
of measurement. Thus according to the present approach,
the collapse of the wave function is not required.


\end{document}